\documentclass{article}

\usepackage{PRIMEarxiv}

\usepackage[utf8]{inputenc} 
\usepackage[T1]{fontenc}    
\usepackage{hyperref}       
\usepackage{url}            
\usepackage{booktabs}       
\usepackage{amsfonts}       
\usepackage{nicefrac}       
\usepackage{microtype}      
\usepackage{lipsum}
\usepackage{fancyhdr}       
\usepackage{graphicx}       
\graphicspath{{media/}}
\usepackage{array}
\usepackage{minted}
\usepackage{listings}
\usepackage{xcolor}
\usepackage{cellspace}
\usepackage{orcidlink}

\hypersetup{
    colorlinks=true,
    linkcolor=black,
    urlcolor=blue,
    citecolor=black
}

\lstdefinestyle{json}{
  basicstyle=\ttfamily\footnotesize,
  numbers=left,
  numberstyle=\tiny\color{gray},
  stepnumber=1,
  numbersep=5pt,
  stringstyle=\color{blue},
  showstringspaces=false,
  breaklines=true,
  frame=single,
  backgroundcolor=\color{gray!5},
}

\title{\textnormal{A Lightweight Hybrid Publish/Subscribe Event Fabric \\for IPC and Modular Distributed Systems}}

\author{
  Dimitris Gkoulis\,\orcidlink{0000-0002-7428-5753} \\
  Harokopio University of Athens \\
  \texttt{gkoulis@hua.gr} \\
}

\begin{document}
\maketitle

\begin{abstract}
Modular software deployed on mini compute units in controlled distributed environments often needs two messaging paths: low-overhead in-process coordination and selective cross-node distribution. In practice, event identity, serialization, and transport bridging are frequently implemented as ad hoc glue, which complicates inter-process communication (IPC), structured routing, and shutdown behavior \cite{eugster2003many,laigner2024empirical,kul2021survey}. This paper presents CNS, a lightweight local-first hybrid event fabric centered on asynchronous fire-and-forget messaging. CNS combines a typed event key, per-family serialization and validation, a local publish/subscribe context for in-process coordination, and a NATS-backed distributed context for inter-node distribution \cite{fernando2021designing,lazidis2022open}. A bridge runtime moves events between the two contexts while preserving a common routing vocabulary. The primary operating model is fire-and-forget publication and subscription; bidirectional request-reply remains available as a secondary extension on the same subject space. A Python prototype and single-machine measurements are reported. Local-only delivery averaged about 30~$\mu$s. Distributed-only delivery averaged 1.26--1.37~ms, and the hybrid bridge averaged 1.64--1.89~ms. Validation introduced modest overhead relative to serialization choice. The resulting artifact is suited to structured IPC and practical message movement within modular services and across bounded sets of controlled nodes.
\end{abstract}

\keywords{event-driven systems \and local-first messaging \and inter-process communication \and NATS \and structured routing \and publish/subscribe \and microservices}

\section{Introduction}
\label{section:introduction}

Modular distributed systems rarely operate in a single communication mode. Software running on mini compute units often requires inexpensive in-process coordination among internal modules and selective distribution of events across process or node boundaries. The first path is primarily an IPC problem. The second introduces transport concerns such as subject routing, serialization, connection management, and shutdown behavior. In many implementations, these paths are developed separately and joined later with custom adapters. The result is usually functional but structurally inconsistent: local coordination follows one model, distributed messaging follows another, and message movement between the two becomes a maintenance burden.

This design pressure is especially visible in regulated and tightly controlled deployments. Examples include industrial controllers, laboratory instruments, validation rigs, and other bounded environments in which each node runs a small modular software stack and external communication is explicit. In such settings, the central requirement is not an open-ended service mesh. It is dependable movement of well-typed signals inside a node and controlled export of selected signals across nodes. Fire-and-forget messaging is therefore the primary semantic requirement. Bidirectional exchanges may be needed in a smaller set of cases, but they do not define the core of the system.

Recent empirical work on event-driven microservices reports persistent difficulties around event schema design, payload management, debugging, auditing, and ordering constraints \cite{laigner2024empirical,lakhai2024method}. These difficulties appear even when the underlying broker is mature. The problem is often not the transport alone. It is the lack of a coherent software structure around event identity, schema handling, and the boundary between local and distributed execution.

CNS addresses that problem as a software artifact rather than as a new broker. The artifact combines a typed event taxonomy, a local publish/subscribe context for in-process coordination, a NATS-backed distributed context for inter-node messaging, and an explicit bridge that transfers events between them. Local coordination is treated as a first-class execution path, not as a reduced case of remote messaging. This orientation is useful when a single service is internally composite, for example when acquisition, normalization, policy, and persistence modules exchange events on one node while selected updates are exported to peer nodes.

The intended scope is deliberately bounded. CNS is designed for modular distributed systems implemented within a single project boundary, deployed on controlled nodes, and organized around structured fire-and-forget event movement. Within that scope, the artifact aims to make routing, validation, and transport bridging explicit and reusable. Request-reply remains available on the same subject space when a bounded bidirectional exchange is required, but it is secondary to the publish/subscribe core.

These concerns sit within a broader line of work on publish/subscribe and event-based middleware. Foundational work described distributed asynchronous collections, type-based interaction, and the core dimensions of time, space, and synchronization decoupling \cite{eugster2000distributed,eugster2000type,eugster2007type,eugster2003many}. Broader surveys and reviews have covered publish/subscribe middleware, message-oriented middleware, and event-driven architectural patterns \cite{liu2003survey,yusuf2004survey,tarkoma2012publish,yongguo2019message,thondalapally2025event}. In the microservice setting, architectural context, communication choices, design trade-offs, maintainability, and event-management difficulties remain active concerns \cite{cerny2018contextual,kul2021survey,vale2022designing,vsandor2024designing,lakhai2024method,laigner2024empirical,elmdahl2023investigation}. Quality of service, self-management, experimental performance, and reliability have likewise remained central evaluation dimensions in publish/subscribe systems \cite{bellavista2014quality,behnel2006quality,jaeger2007self,rizano2013experimental,esposito2012tutorial}. Comparative and domain-oriented studies further motivate bounded deployment-aware designs for edge, IIoT, and application-specific event fabrics \cite{lazidis2022publish,lazidis2022open,nast2022survey,rausch2017message,pellegrino2014pushing,roffia2016semantic,perera2022common,9497471,fi15040135,gkoulis2026selfhealing,korhonen2025centering}.

The paper makes three concrete contributions. First, it defines a typed event-routing model in which structured event keys serve simultaneously as identifiers, routing subjects, and schema anchors. Second, it presents a hybrid runtime that bridges local publish/subscribe coordination with distributed asynchronous messaging over NATS. Third, it reports prototype measurements that characterize latency, throughput, serialization and validation overhead, and shutdown behavior for the current implementation. The emphasis throughout is operational clarity and bounded applicability rather than broad claims of general-purpose middleware.

The remainder of the paper is organized as follows. Section~\ref{section:context} states the design goals and scope. Section~\ref{section:architecture} presents the architecture. Section~\ref{sec:implementation} describes the prototype implementation. Section~\ref{sec:methodology} summarizes the evaluation methodology. Section~\ref{sec:results} reports the results. Section~\ref{sec:applicability} defines the intended applicability of the artifact. Section~\ref{sec:limitations} states actionable limitations and future work. Section~\ref{section:conclusions} concludes.

\section{Design Goals and Scope}
\label{section:context}

CNS was designed around six goals.

First, event identity should be an explicit architectural element. Topic names and subjects often become opaque strings that gradually accumulate inconsistent conventions. CNS instead uses a structured event key whose components define the routing taxonomy directly.

Second, serialization and validation should be registered per event family rather than distributed across producers and consumers. This keeps schema handling close to the event definition and reduces transport-specific branching in application code.

Third, local coordination should remain explicit and inexpensive. Inside one process or one tightly coupled service, modules should be able to exchange events without incurring the full machinery of distributed messaging. In CNS, the local context is therefore a primary runtime layer and not a fallback mode.

Fourth, distributed interoperability should rely on a transport that already supports hierarchical subjects, wildcard subscriptions, and request-reply on the same routing substrate. NATS satisfies those requirements and aligns naturally with the structured event taxonomy used here \cite{fernando2021designing,lazidis2022open}.

Fifth, movement between local and distributed contexts should be visible in the design. CNS does not hide both paths behind a single abstract bus. It uses explicit transfer loops so that message movement across the process boundary remains inspectable and measurable.

Sixth, the artifact should specialize cleanly. Project-specific event families and qualifiers should be introduced through registries and context configuration, not by rewriting the transport core.

These goals establish the intended operating profile of the artifact. CNS is designed for bounded deployments in which nodes are known, message families are controlled, and software modules are developed together. The dominant semantic model is asynchronous fire-and-forget delivery. Bidirectional communication is supported when correlated request and response pairs are necessary, but it is not the organizing principle of the system.

\section{Architecture}
\label{section:architecture}

\begin{figure}[H]
  \centering
  \includegraphics[width=0.9\textwidth]{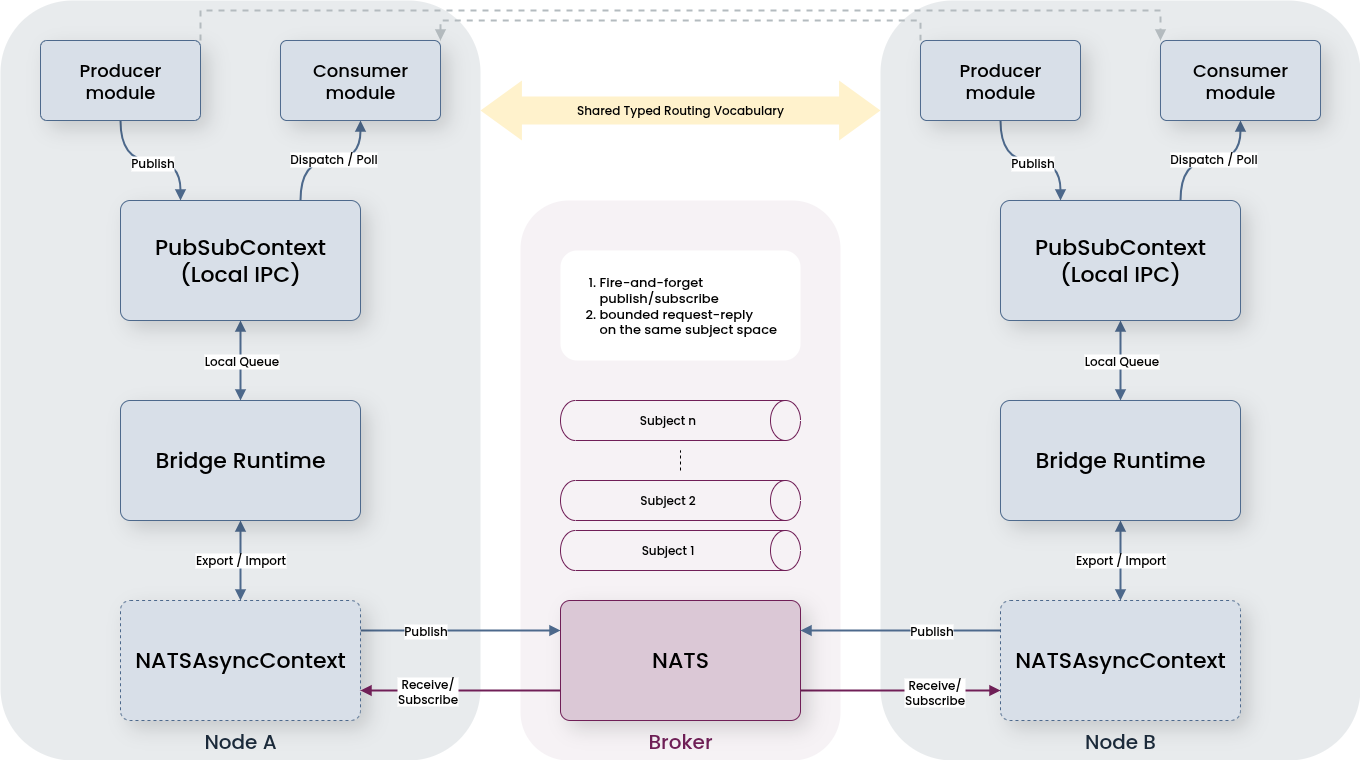}
  \caption{Local-first hybrid event fabric spanning local IPC and distributed transport. Each node contains a local publish/subscribe context, a bridge runtime, and a NATS-backed asynchronous context. Events share one typed routing vocabulary across local and distributed paths.}
  \label{fig:architecture}
\end{figure}

\subsection{Event Model}
\label{subsection:eventmodel}

At the center of CNS is the \texttt{EventTypeKey}, a structured identifier with the components \texttt{space}, \texttt{super\_family}, \texttt{family}, \texttt{name}, and \texttt{qualifiers}. The full key is assembled as a dot-separated string. This design mirrors subject-based brokers such as NATS, where subjects are hierarchical, dot-separated, and matched through wildcard subscriptions \cite{fernando2021designing}.

The key also exposes several derived forms. The \texttt{group\_key} identifies a broader cluster of related events. The \texttt{base\_key} identifies the event family independently of qualifiers. The \texttt{qualifiers\_key} captures the context-dependent suffix, such as a node identifier, module identifier, or sampling window. An example in the intended deployment style is \texttt{fabric.node.status.update.node17.10s}, where the qualifier sequence binds a generic status family to a specific node and interval.

An \texttt{Event} consists of the structured key, a payload, and a metadata dictionary. Payload typing is handled through an \texttt{EventTypeSerDe} object registered against the \texttt{base\_key}. That object bundles serialization, deserialization, and validation. This keeps the transport layer generic while making event-family handling explicit. When a new family is introduced, the implementation registers one serializer, one deserializer, and an optional validator, rather than editing multiple transport paths.

This organization keeps the event family, rather than the producing module, as the primary schema anchor. It also aligns routing and data handling: the same base key that selects subscribers also selects the SerDe and validation logic. That alignment is central to the architecture.

\subsection{Local publish/subscribe context}

The local context is represented by \texttt{PubSubContext}, a small in-process publish/subscribe mechanism with two queues: one queue for outgoing publish requests and one queue for subscribed events to be polled or consumed. Local subscriptions are stored in a map from pattern to handler. When a subscription is declared without a handler, matched events are placed in the local subscription queue. When a handler is present, it is invoked directly.

This dual mode supports two common uses inside a modular node. The first is direct local dispatch for lightweight callbacks. The second is staged processing, where internal modules prefer to drain events explicitly from a queue. Both patterns are useful in IPC-heavy software, and both remain within the same routing vocabulary.

Pattern matching is wildcard-aware and aligned with NATS subject semantics \cite{fernando2021designing}. Structured in-process event movement and publisher/subscriber IPC have also been studied in real-time settings, which reinforces the choice to keep the local path explicit rather than treating it as a broker-only special case \cite{rajkumar1995real,rajkumar1996high,park2026efficient}. Exact tokens, single-token wildcards, and suffix wildcards follow the same logic in local and distributed paths. As a result, a module can subscribe to a family of local events and later expose the same family to the distributed context without changing its routing conventions.

The local context is intentionally lightweight. It does not provide durable storage, complex broker semantics, or broad delivery guarantees. Its role is structured IPC inside one runtime boundary and one endpoint of the bridge.

\subsection{Distributed asynchronous context}

The distributed context is implemented by \texttt{NATSAsyncContext}, which wraps a NATS client adapter. NATS provides hierarchical subjects, wildcard subscriptions, and request-reply over the same subject space \cite{fernando2021designing,lazidis2022open}. These properties fit the routing model of CNS directly.

On publication, the distributed context resolves the appropriate \texttt{EventTypeSerDe} from the event's \texttt{base\_key}. Validation is applied when configured. Serialization then uses the registered serializer or, in the current prototype, a default Python mechanism when no custom serializer is registered. Metadata headers are enriched with the full event key, the base key, the qualifiers key, timing information, and client identity fields. This metadata allows the receiving side to reconstruct the event without inferring payload type from the message body alone.

On consumption, the distributed context inspects headers, reconstructs the structured key, resolves the relevant deserializer and validator, and produces a local \texttt{Event} instance. Messages that cannot be reconstructed or validated are rejected with warnings. This makes schema failure visible at the context boundary rather than allowing silent corruption.

Metadata propagation is implemented through NATS headers. CNS uses an application-specific header prefix for runtime metadata so that transport metadata and application metadata remain distinct while preserving transport-native header handling.

\subsection{Bridge runtime}

The bridge between local and distributed contexts is the central runtime mechanism of CNS. It is implemented through \texttt{IPrivateSharedLocalContext}, which coordinates a local \texttt{PubSubContext}, a distributed \texttt{NATSAsyncContext}, an event loop, a worker thread, and a thread pool for selected handoffs.

Two transfer loops define the bridge semantics. The first moves events from the local publish queue to the distributed publisher. This allows synchronous local producers to emit events without depending directly on the asynchronous network API. The second moves events from the distributed subscription queue into the local subscription context. Remote messages are thereby localized and reintroduced into the same pattern-matching system used by internal modules.

The result is explicit message movement across local and distributed contexts. A module can publish locally, another node can receive the distributed form, and the destination node can localize the event again for its own internal modules. For example, a local acquisition module may publish a status update, the bridge may export that update to peer nodes, and a remote recorder module may consume the localized copy without depending on transport-specific code.

The current prototype runs the asynchronous event loop in a dedicated thread. This isolates network communication from the main synchronous application path while preserving an asynchronous transport implementation. A thread pool is used where blocking local operations must not stall the event loop. This design is adequate for the prototype and exposes the remaining hardening work clearly: ordering guarantees, queue discipline, and shutdown behavior depend on the exact coordination between these runtime components.

\subsection{Bounded bidirectional communication}

The primary CNS semantics are publish/subscribe and fire-and-forget delivery. Bidirectional communication is supported on the same subject space when a bounded request and correlated response are required between controlled peers. NATS request-reply uses ordinary subjects together with a reply subject \cite{fernando2021designing}, which means CNS can support request and response flows without introducing a separate RPC substrate.

In practice, request subjects and reply subjects can be modeled either through transport-level request-reply or through explicit request and response event families. The same registry mechanism used for ordinary events can therefore validate and deserialize bidirectional messages. This keeps routing and schema handling uniform.

The bidirectional path remains secondary in the present design. It is suitable for cases such as a node querying a peer for a short-lived status snapshot or acknowledging a control step, but the artifact is not organized around general-purpose service invocation. Correlation metadata, timeout policy, and typed convenience helpers are therefore identified as follow-on engineering work rather than as the center of the current system.

\subsection{Specialization pattern}

The generic runtime is intended to support project-specific event registries without transport modification. A specialization defines a registry of event families, each with its payload expectations and validator, and then constructs full event keys by combining those families with runtime qualifiers.

In the intended scope, a project may define families for status updates, control acknowledgements, snapshots, synchronization signals, or health events. Qualifiers can then bind those families to concrete node identifiers, module identifiers, or time windows. This approach scales the event taxonomy without changing bridge logic or transport logic. The transport core remains stable, while the project-level routing vocabulary is expressed through registries and configuration.

\section{Prototype Implementation}
\label{sec:implementation}

The prototype is implemented in Python using \texttt{asyncio}, standard queue abstractions, thread-based concurrency for bridge execution, and a NATS client library for distributed transport. NATS provides asynchronous publish/subscribe support in Python, which matches the implementation strategy adopted here \cite{fernando2021designing,lazidis2022open}.

The software is organized into four layers. The first layer contains the generic data types and helpers, including event keys, events, wildcard subject matching, and SerDe registration structures. The second layer implements local publish/subscribe behavior. The third layer implements the NATS-backed distributed context. The fourth layer contains the bridge interface together with project-specific specializations. In normal extension work, new event definitions and registries should be added at the specialization layer rather than by modifying the transport core.

Several implementation choices are central to the artifact. The first is explicit validation at the event-family boundary. This is lighter than a separate schema registry and catches mismatches close to the transport boundary. The second is header enrichment. Event identity is carried both in the subject and in transport headers, which simplifies reconstruction and diagnostics on receipt. The third is support for both queue-based and callback-based local subscriptions. This accommodates both staged internal pipelines and direct callback dispatch inside one node.

The fourth implementation choice is explicit stop handling. Bridge loops are controlled by an active-state flag, and the NATS adapter attempts to drain and unsubscribe subscriptions before closing the connection. The current implementation already treats shutdown as part of the runtime contract, even though the reported results show that backlog handling requires further engineering. This is appropriate for an artifact centered on message movement across local and distributed contexts.

A complete artifact package for this prototype would pair the code with runnable examples, API-level tests, and the benchmark harness.

\section{Evaluation Methodology}
\label{sec:methodology}

The evaluation examines four runtime questions.

The first question concerns latency overhead across the three communication paths exposed by the architecture: local-only publication, distributed-only publication, and local-to-distributed-to-local transfer through the hybrid bridge. The second concerns throughput across the same paths. The third isolates the effect of serialization and validation choices in the distributed path. The fourth examines graceful-stop behavior under forced shutdown while backlog remains in flight.

The benchmark environment is a single-machine setup and should be interpreted accordingly. For each payload size, namely 256~B, 1~KiB, and 4~KiB, the reported measurements use three repetitions after one warm-up run. The latency benchmark uses 3{,}000 messages per run. The throughput benchmark uses 50{,}000 messages per run. The graceful-stop benchmark attempts 100{,}000 messages before forcing shutdown after 2.0~s.

The baseline comparisons are architectural rather than broker-to-broker. Local-only measurements isolate the in-process IPC path. Distributed-only measurements isolate broker-backed message movement without local bridging. Hybrid measurements characterize the additional cost of moving an event from local publication, through the distributed context, and back into a local subscription context on the receiving side. This isolates the cost of the bridge itself.

The reported study characterizes the current prototype on a laptop-class host. It is sufficient to identify the performance hierarchy of the three paths, estimate the overhead of validation and serialization choices, and expose shutdown behavior under load. This measurement profile is consistent with prior publish/subscribe work that isolates latency, throughput, and reliability behavior before broader deployment studies \cite{rizano2013experimental,bellavista2014quality,esposito2012tutorial}. Multi-host experiments and maintainability measurements remain outside the present results and are treated as future work in Section~\ref{sec:limitations}.

\section{Results}
\label{sec:results}

\subsection{Experimental setting}
\label{subsec:results-setting}

The full benchmark preset exceeded the capabilities of the target computing unit, so the reported measurements use the light configuration described in Section~\ref{sec:methodology}. The results therefore characterize a bounded single-machine deployment rather than a large-scale stress test. This matches the current artifact stage and the intended node-level operating profile.

\subsection{End-to-end latency}
\label{subsec:results-latency}

Table~\ref{tab:q1-latency} summarizes latency across the three communication paths. The local-only path is consistently the fastest. Mean latency remains close to 30~$\mu$s across all payload sizes, with \(p95\) below 54~$\mu$s and \(p99\) below 98~$\mu$s. For the intended IPC role inside one node, the local context therefore adds little per-message cost under the measured workload.

The distributed-only path raises mean latency to 1.26--1.37~ms, and the hybrid bridge raises it further to 1.64--1.89~ms. Relative to local-only delivery, distributed-only delivery is roughly 42--46 times slower in mean latency, while the hybrid bridge is roughly 55--64 times slower. The dominant step change comes from crossing the process and broker boundary. The bridge adds a further but smaller increment on top of that transition.

Across payload sizes, hybrid mean latency is about 1.31 times the distributed-only mean at 256~B, about 1.31 times at 1~KiB, and about 1.38 times at 4~KiB. This indicates that the bridge cost is measurable but remains below the main cost of distributed transport itself.

Tail latency is more variable than the averages. For the distributed-only path, mean \(p99\) latency ranges from about 2.34~ms to 2.99~ms. For the hybrid path, mean \(p99\) latency ranges from about 3.48~ms to 4.85~ms. Individual maxima are higher, reaching 12--18~ms in some runs. On the measured host, tail behavior is therefore influenced by scheduling and transient contention in addition to runtime structure.

\begin{table}[t]
\centering
\caption{Mean latency across three repetitions. Values are reported as mean $\pm$ standard deviation.}
\label{tab:q1-latency}
\begin{tabular}{llccc}
\toprule
Payload & Path & Mean latency ($\mu$s) & $p95$ ($\mu$s) & $p99$ ($\mu$s) \\
\midrule
256~B & local-only & $30.1 \pm 3.8$ & 53.9 & 97.9 \\
256~B & distributed-only & $1257.2 \pm 141.9$ & 1848.2 & 2992.6 \\
256~B & hybrid bridge & $1642.4 \pm 90.6$ & 2227.2 & 3477.5 \\
\midrule
1~KiB & local-only & $30.3 \pm 4.3$ & 46.0 & 68.9 \\
1~KiB & distributed-only & $1369.8 \pm 78.8$ & 1751.8 & 2339.2 \\
1~KiB & hybrid bridge & $1792.8 \pm 120.6$ & 2236.4 & 3868.9 \\
\midrule
4~KiB & local-only & $29.6 \pm 3.8$ & 50.6 & 95.4 \\
4~KiB & distributed-only & $1364.2 \pm 31.0$ & 1911.2 & 2603.3 \\
4~KiB & hybrid bridge & $1888.0 \pm 153.2$ & 2419.4 & 4846.6 \\
\bottomrule
\end{tabular}
\end{table}

\subsection{Throughput}
\label{subsec:results-throughput}

Table~\ref{tab:q2-throughput} reports throughput. Local-only delivery achieves the highest rate, ranging from about 34.4k to 39.7k messages/s. Distributed-only throughput ranges from about 4.56k to 6.76k messages/s. The hybrid bridge remains near 1.19k to 1.23k messages/s across the measured payload sizes.

The hybrid path varies only slightly with payload size, which suggests that the main bottleneck in the current prototype is bridge coordination rather than payload serialization cost alone. Queue handoff, interaction between the synchronous and asynchronous sides, and event-loop coordination are likely to dominate message-size effects at this scale.

Local-only throughput is about 29--32 times higher than hybrid throughput and about 5--8 times higher than distributed-only throughput. Distributed-only throughput is about 3.7--5.5 times higher than hybrid throughput. These ratios locate the present cost of the bridge clearly. The structured routing model is operationally viable, but the bridge does not provide its functionality at negligible cost.

\begin{table}[t]
\centering
\caption{Throughput across three repetitions. Values are reported as mean $\pm$ standard deviation.}
\label{tab:q2-throughput}
\begin{tabular}{llc}
\toprule
Payload & Path & Throughput (messages/s) \\
\midrule
256~B & local-only & $37877 \pm 4687$ \\
256~B & distributed-only & $6759 \pm 517$ \\
256~B & hybrid bridge & $1226 \pm 40$ \\
\midrule
1~KiB & local-only & $34418 \pm 6569$ \\
1~KiB & distributed-only & $5765 \pm 209$ \\
1~KiB & hybrid bridge & $1192 \pm 52$ \\
\midrule
4~KiB & local-only & $39676 \pm 1408$ \\
4~KiB & distributed-only & $4564 \pm 180$ \\
4~KiB & hybrid bridge & $1225 \pm 14$ \\
\bottomrule
\end{tabular}
\end{table}

\subsection{Validation and serialization overhead}
\label{subsec:results-serde}

The third measurement series examines the distributed path under three cases: pickle with validation, pickle without validation, and JSON with validation.

Validation itself adds limited overhead in the measured configuration. At 256~B, validation on top of pickle changes mean latency negligibly and reduces throughput by about 4.1\%. At 1~KiB, latency remains nearly unchanged while throughput falls by about 11.8\%. At 4~KiB, the measurements are noisier and do not show a consistent penalty. Across the measured runs, validation is therefore not the main cost center.

Serialization choice has a more visible effect. JSON with validation generally shows higher mean latency than pickle with validation, especially at 256~B and 4~KiB, and lower throughput at those payload sizes. At 256~B, JSON with validation reduces throughput by about 10.0\% relative to pickle with validation. At 4~KiB, the reduction is about 16.1\%. The 1~KiB case is close to neutral, which is consistent with the limited scale of the benchmark.

These results support retaining per-family validation in the architecture. Within the measured workload, serialization choice has more influence on throughput than the validator hook itself.

\begin{table}[t]
\centering
\caption{Distributed-path validation/serialization comparison. Values are reported as mean $\pm$ standard deviation.}
\label{tab:q3-serde}
\begin{tabular}{llcc}
\toprule
Payload & Case & Mean latency ($\mu$s) & Throughput (messages/s) \\
\midrule
256~B & pickle, no validation & $1175.9 \pm 65.0$ & $7244 \pm 461$ \\
256~B & pickle, with validation & $1176.3 \pm 58.5$ & $6947 \pm 365$ \\
256~B & JSON, with validation & $1223.2 \pm 105.4$ & $6255 \pm 377$ \\
\midrule
1~KiB & pickle, no validation & $1266.8 \pm 140.0$ & $6597 \pm 508$ \\
1~KiB & pickle, with validation & $1260.9 \pm 138.2$ & $5818 \pm 639$ \\
1~KiB & JSON, with validation & $1280.2 \pm 127.3$ & $5934 \pm 263$ \\
\midrule
4~KiB & pickle, no validation & $1320.1 \pm 165.9$ & $4821 \pm 254$ \\
4~KiB & pickle, with validation & $1239.6 \pm 7.2$ & $4990 \pm 299$ \\
4~KiB & JSON, with validation & $1354.5 \pm 14.6$ & $4187 \pm 86$ \\
\bottomrule
\end{tabular}
\end{table}

\subsection{Graceful-stop behavior}
\label{subsec:results-stop}

Table~\ref{tab:q4-stop} reports graceful-stop behavior for the hybrid bridge under forced shutdown. Completion remains low: the bridge completes only about 1.19\% to 1.25\% of messages before shutdown, corresponding to roughly 1{,}189 to 1{,}253 messages received out of 100{,}000 sent before stop. Estimated loss ranges from approximately 98{,}747 to 98{,}811 messages.

This result identifies a specific weakness in the current prototype. Under sustained load, once the stop signal arrives with substantial backlog still in flight, the present bridge does not preserve most queued work. The test is intentionally harsh: the producer is stopped after 2.0~s, the active-state flag is cleared, the system waits briefly, and the worker thread is then joined. Even under that condition, the completion rate is too low to support stronger shutdown claims.

The result is operationally important because the architecture already treats stop behavior as part of the runtime contract. Queue discipline, bounded backlog, and drain semantics therefore require further engineering before the artifact can support stronger delivery claims under forced termination.

\begin{table}[t]
\centering
\caption{Graceful-stop results for the hybrid bridge. Values are reported as mean $\pm$ standard deviation.}
\label{tab:q4-stop}
\begin{tabular}{lccc}
\toprule
Payload & Completion rate & Received before join & Estimated lost \\
\midrule
256~B & $0.01253 \pm 0.00030$ & $1253 \pm 30$ & $98747 \pm 30$ \\
1~KiB & $0.01224 \pm 0.00079$ & $1224 \pm 79$ & $98776 \pm 79$ \\
4~KiB & $0.01189 \pm 0.00043$ & $1189 \pm 43$ & $98811 \pm 43$ \\
\bottomrule
\end{tabular}
\end{table}

\subsection{Discussion}
\label{subsec:results-discussion}

The measurements present a consistent picture of the current artifact. The latency results preserve the expected hierarchy: local-only communication is fastest, distributed-only communication is slower because it crosses the broker boundary, and the hybrid bridge adds an additional but smaller cost. Throughput results show the same structure in inverse form and locate bridge coordination as a principal throughput constraint. Validation contributes only a modest share of the distributed-path cost, while serialization format has a somewhat larger effect. Graceful-stop behavior under backlog is the weakest measured aspect of the prototype.

The present evaluation therefore supports a bounded claim. CNS provides a measurable and operationally legible architecture for structured fire-and-forget event movement across local and distributed contexts. The prototype already supports the intended IPC role inside one node and selective cross-node distribution. The current engineering priorities are bridge throughput, tail-latency stabilization, and especially stop semantics under backlog.

\section{Intended Applicability}
\label{sec:applicability}

CNS is intended for modular distributed systems built and operated within a controlled project boundary. The typical deployment consists of mini compute units that run several cooperating modules locally and exchange selected event families with peer nodes. In this setting, the local context serves as structured IPC inside one node, while the distributed context carries a subset of events across nodes under the same routing vocabulary.

A representative use pattern is a node that contains acquisition, normalization, policy, and recorder modules. These modules exchange local events through the \texttt{PubSubContext}. Selected families, such as status updates, synchronization pulses, control acknowledgements, or state snapshots, are then exported through the bridge to the NATS subject space. On a receiving node, the distributed message is reconstructed and localized again so that internal modules can consume it without transport-specific code.

This operating style fits controlled environments in which routing conventions, message families, and node identities are known in advance. It also fits software developed as modular parts inside a single project, where distribution ergonomics matter but a full broker-centric platform would add unnecessary weight. Related event-driven systems and domain specializations have appeared in semantic IoT, event-cloud, edge-computing, smart-farming, and finance-oriented settings \cite{roffia2016semantic,pellegrino2014pushing,perera2022common,9497471,fi15040135,gkoulis2026selfhealing,korhonen2025centering}. The design is therefore appropriate for structured message movement across local and distributed contexts when fire-and-forget delivery is primary and bidirectional exchanges are occasional and bounded.

\section{Limitations and Future Work}
\label{sec:limitations}

Several improvements follow directly from the present prototype and measurements.

First, the ordering and thread-safety semantics of the local context should be specified more strictly and enforced explicitly. The current architecture makes these questions visible, but the guarantees should be stated with greater precision.

Second, queue discipline, backpressure, and draining policy require additional runtime design. The graceful-stop measurements show that forced termination under backlog is the main unresolved operational problem in the prototype.

Third, the default serialization path is convenient for prototyping but insufficient for broader interoperability. A language-neutral serializer should be provided as a first-class option together with explicit compatibility expectations.

Fourth, the bounded request-reply path would benefit from typed helpers for correlation identifiers, timeout handling, and reply-family declaration. The current architecture supports such flows conceptually, but the convenience layer remains incomplete.

Fifth, the evaluation should be extended beyond the present single-machine configuration. Multi-host experiments, broader workload variation, and maintainability measurements for adding new event families would provide a fuller characterization of the artifact.

\section{Conclusions}
\label{section:conclusions}

CNS is a lightweight local-first hybrid event fabric for modular distributed systems that need structured fire-and-forget message movement across local and distributed contexts. The artifact combines a typed event taxonomy, per-family serialization and validation, a local publish/subscribe context for IPC, a NATS-backed distributed context, and an explicit bridge between them. Within its intended scope, this design keeps routing and schema handling explicit while allowing local modules and distributed peers to share one subject vocabulary.

The reported measurements show a clear runtime hierarchy. Local-only communication remains fast, distributed communication introduces the main latency and throughput step change, and the hybrid bridge adds a further measurable cost. Validation overhead is modest relative to serialization choice. Graceful-stop behavior under backlog is the main unresolved engineering issue in the current prototype.

The artifact is suited to modular software stacks running on mini compute units in controlled distributed environments, especially where structured IPC and practical cross-node distribution are required inside one project boundary. Further work should tighten queue semantics, stop behavior, interoperability, and multi-host evaluation without changing the core fire-and-forget architecture.

\section*{Code and Data Availability}

The source code for the prototype implementation is publicly available at
\url{https://github.com/gkoulis/nats-cns/tree/main/src/nats_cns}.

The dataset used in this study is publicly available at
\url{https://github.com/gkoulis/nats-cns/tree/main/benchmark_results_paper}.

The code used for the experimental evaluation and reproduction of the reported results is publicly available at
\url{https://github.com/gkoulis/nats-cns/tree/main/benchmarks}.

\bibliographystyle{unsrt}
\bibliography{references}

\end{document}